\documentclass[iop]{emulateapj}
\def\am {Am$^{241}$}  
\def\lax {\ifmmode{_<\atop^{\sim}}\else{${_<\atop^{\sim}}$}\fi}  
\def\gax {\ifmmode{_>\atop^{\sim}}\else{${_>\atop^{\sim}}$}\fi}  
\def\gtorder{\mathrel{\raise.3ex\hbox{$>$}\mkern-14mu
             \lower0.6ex\hbox{$\sim$}}}

\def\xte{{\it RXTE}}

\begin{document}

\title{ Advances in the {\it RXTE} Proportional Counter Array Calibration: Nearing the Statistical  Limit}

\author{Nikolai Shaposhnikov\altaffilmark{1,2}, Keith Jahoda\altaffilmark{2}, Craig Markwardt\altaffilmark{2}, Jean Swank\altaffilmark{2}, Tod Strohmayer\altaffilmark{2}}

\altaffiltext{1}{CRESST/University of Maryland, Department of Astronomy, College Park MD, 20742, nikolai.v.shaposhnikov@nasa.gov}

\altaffiltext{2}{Goddard Space Flight Center, NASA, Astrophysics Science Division, Greenbelt MD 20771}

\altaffiltext{3}{CRESST/Universities Space Research Association, Columbia MD, 21044}

\begin{abstract}
During its 16 years of service the Rossi X-ray Timing Explorer (\xte )  mission has provided an extensive  archive of data, which will serve as a primary source of high cadence observations of variable X-ray sources for fast timing studies. It is, therefore, very important to have the most reliable calibration of \xte\ instruments.   The Proportional Counter Array (PCA) is the primary instrument on-board \xte\ which provides data in 3-50 keV energy range with sub-millisecond time resolution 
in up to 256 energy channels. 
In 2009 the RXTE team revised the response residual minimization method used to derive the parameters of the 
PCA physical model. The procedure is based on the residual minimization between the  model spectrum for Crab
nebula emission and a calibration data set
consisting of a number of spectra from the Crab and the on-board Am$_{241}$ calibration source, uniformly
covering the whole \xte\ mission operation period. The new method led to a much more effective model convergence and  allowed for better understanding of the PCA energy-to-channel relationship. It greatly improved the response matrix performance. We describe the new version of the \xte/PCA response generator PCARMF v11.7 (HEASOFT Release 6.7)along with the corresponding energy-to-channel conversion table (verson {\it e05v04}) and their difference from the previous releases of PCA calibration. The new PCA response adequately represents the spectrum of the calibration sources and successfully predicts the energy of the narrow iron emission line in Cas-A throughout the RXTE mission. 
\end{abstract}


\section{Introduction}
The Rossi X-ray Timing Explorer (\xte) was launched on December 30, 1995 and successfully operated until January 4, 2012.  \xte\ is an X-ray observatory with a powerful and unique combination of large collecting area, broad-band spectral coverage, high time resolution, flexible scheduling, and ability of quick response and frequent monitoring of time-critical targets of opportunity. \xte\ observations have led to breakthroughs in our understanding of physics of strong gravity, high density, and intense magnetic field environments found in neutron stars, galactic and extragalactic black holes and other sources. The mission combined two pointing instruments, the Proportional Counter Array (PCA) developed to provide data for energies between 3 and 50 keV, 
and the High Energy X-ray Timing Experiment \citep[HEXTE;][]{hexte}  covering the 20-250 keV energy range. These instruments 
were equipped with collimators yielding a FWHM of one degree. In addition, RXTE carried an All-Sky Monitor \citep[ASM;][]{asm}
that scans about 80\% of the sky every orbit, allowing monitoring at time scales of 90 minutes or longer. 
Data from PCA and ASM are processed on board by the Experiment Data System (EDS). 

The PCA is array of five large-area proportional counter units (PCUs) designed to perform observations of bright X-ray sources with high timing and modest spectral resolution. The main chamber of each PCU is divided into three volumes or layers filed with xenon. In addition, all PCUs were initially equipped with a propane-filled '"veto" layer in front of 
the top xenon layer.  The calibration of the PCA, as well as the details of its design and operation,
are described in \citet[][J06 hereafter]{jah06}. The response generation software for the PCA is based on physical model of the instrument. The main components of the model are the quantum efficiency, which gives the probability of an
 X-ray  photon to be absorbed in one of the detector volumes, and the redistribution matrix, which provides the probability for a photon to be detected in one of the PCU energy channels.  The  model has a complex dependence on
 many parameters, which have to be properly optimized to minimize
a difference between the predicted model and the observed spectrum of one or more 
calibration sources, i.e. sources with well known spectral characteristics. Implementation of 
an effective parameter optimization procedure is vital in performing this task.

The set of PCA  parameters describing the instrument response since the start of the mission and until 2004 has been calculated in J06. However, calibration observations of the Crab and other sources after 2004 suggested that the model and its parameters have to be updated to provide a consistent response for new science observations. 
In 2009 we revised the PCA model and the response  minimization method. 
The new model provided  significant response improvement for the entire mission span, and especially
 for the data collected after 2004.

In this paper we describe in detail the new fitting procedure and the results of the
PCA response modeling. The paper is structured as follows. In the next section we provide a brief review of 
the PCA physical model and the response generation software. The calibration data is described in \S 3. In \S \ref{setup}
we provide the details of our PCA model implementation and the response minimization fitting environment in XSPEC astrophysical fitting package. We describe and discuss the results in \S \ref{disc}. Conclusions follow in \S
\ref{summary}. 

\section{The PCA detector model and the response generator PCARMF}
\label{pcamodel}

For spectral analysis of PCA data with spectral modeling tools like XSPEC \citep{xspec}, one must supply a response matrix. 
This matrix describes the probability of a photon of a particular energy to be detected in a specific channel of a instrument
detector. Response calculation for PCA is provided by the {\tt pcarmf} tool, which is  a part of FTOOLS astrophysical data analysis environment, maintained by HEASARC\footnote{http://heasarc.gsfc.nasa.gov/docs/software.html}. 

The {\tt pcarmf} tool has several major components: quantum efficiency (i.e. effective area), redistribution matrix (i.e. the spectral resolution), and the energy-to-channel relationship (i.e. the gain). Different components of the response are controlled by various parameters. The quantum efficiency and redistribution parameters are stored in the task parameter file {\tt pcarmf.par}. The energy-to-channel relationship is described by coefficients in a table which resides in calibration database CALDB, the so-called "e2c" file (which can alternatively can be supplied as a stand-alone FITS file). The previous e2c relationship (released in 2004,J06) is referred to as {\it e05v03}, while a new e2c relationship described here  is designated as {\it e05v04}. 


The detailed description of the physical model of the PCA response is presented in J06.
While the major design of the physical model is kept the same in the new response,
some modifications are introduced to improve the model performance. 
During the process of response modeling to obtain a new PCARMF parameter set which best reproduces the calibration data, we have
made necessary additions and modifications to the quantum efficiency and energy-to-channel models.
We provide detailed description of differences between new and previous 
models below.

\begin{figure}
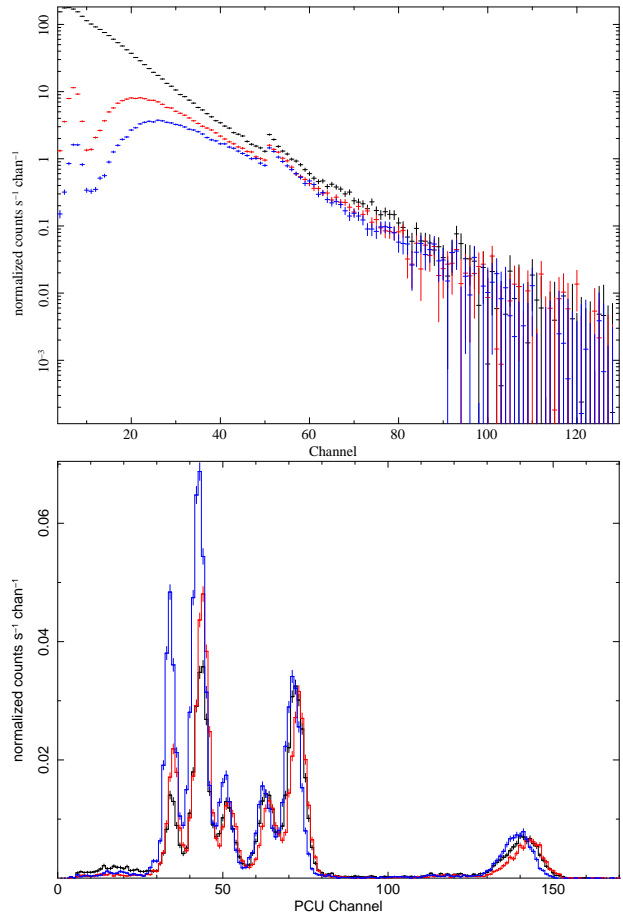

\includegraphics[scale=0.35,angle=-90]{f1a.eps}
\includegraphics[scale=0.35,angle=-90]{f1b.eps}
\caption{ {\it RXTE} calibration data.{\it Top}: Example of the Crab spectra for each layer (black for the top, red for the second and blue for the third layers). {\it Bottom}: The same as the top diagram for the Am$_{241}$ on-board source.}
\label{data}
\end{figure}

The PCA response physical model defines a secondary energy scale $E_p$ of the incoming photon, which
is proportional to the number of electrons produced in the detector. It is related to the actual photon energy
$E_\gamma$ by a factor close to unity, experiencing small variations with energy (see J06 and references therefin). 
The relationship between $E_\gamma$ and $E_p$ is adopted to the new PCA response without any changes.
The new PCA e2c {\it e05v04} now has an "instantaneous" quadratic relationship between channel and $E_p$ (see J06):
\begin{equation}
ch(E,T)=A+BE_p+CE^2_p+DE_p^3,
\label{c2e}
\end{equation}
where 
\begin{equation}
A=A_0+A_1\Delta T 
\label{a}
\end{equation}
and 
\begin{equation}
B=B_0+B_1\Delta T,
\label{b}
\end{equation}
where $\Delta T$ is time in days between 
the observation start time and a reference time
$T_0$, which is taken as December 20, 1997 00:00 UTC (50802.0 MJD).
The above new e2c relationship has several
major differences with respect to the previous response version.
Namely, we drop  the quadratic terms in time dependence of $A$ and $B$, and we add a cubic term in
the energy dependence  (compare with Eq. 3-5 in J06).The energy-to-channel relationship (i.e. six coefficients $A_0$,$A_1$,$B_0$,$B_1$,$C$ and $D$) is determined by various calibration-type observations.

Time dependent terms in the e2c relationship are due to gradual changes 
in density, pressure and volume of xenon gas in the detector chambers 
caused by slow  interchange of gas between the PCU volumes, chamber 
walls bowing and other effects. While the leak of xenon from the top layer into
the propane "veto" layer seems to be the dominant factor, it is extremely difficult
to account for every such effect in  exact manner. Fortunately, allowing the
linear time dependent trends in the linear and quadratic energy terms individually 
for each layer is sufficient to account for the cumulative effect. 
As a result  the values of the parameters $A_1$ and $B_1$ differ significantly from layer to layer
for a particular PCU. 

In addition, abrupt shifts in the relationship 
are caused by such events as PCU high voltage change or propane layer loss. Planned voltage changes were
made on March 21, 1996, April 15, 1996 and March 22, 1999 (J06).  The propane layers of PCUs 0 and 1 were lost 
on May 12, 2000 and December 25, 2006 correspondingly, due to the sudden loss of a veto layer, presumably, 
because of a micrometeorite impact (we adopt 0-4 PCU numbering convention throughout the Paper).
Due to these events  the entire {\it RXTE} mission span is divided into epochs, each having an
 individual sets of  e2c parameters. The previous e2c {\it e03v04} had 5 epochs. The epochs 1,2,3,4 
were divided by voltage changes, while the PCU0 propane pressure loss event defined the start of epoch 5 (see J06). 
We find that for e2c {\it e05v04} the epoch 5 is required only for PCUs 0 and PCU 1 beginning on the propane layer 
loss event (see below). We, therefore, redefine epoch 5 as starting on veto layer loss event and we effectively
drop epoch 5 for PCUs 2,3 and 4. These changes in e2c relationship are the most important ingredients of
 the response improvement.

Another area where PCARMF v11.7 differs significantly from v11.1 is treatment of xenon L-escape lines. In the previous response version L-escape lines were ignored. While L-escape contribution for the PCU layer 1 is indeed negligible, this is not true for layers 2 and 3. Most probably, this can be explained by the following scenario. Almost all photons entering the detector with energies near the L-edge ($\sim$5 keV) are absorbed in the top layer  and most of L-escape photons produced in this layer are vetoed. However, a small fraction of L-escape photons in layer 1 do not get absorbed in the same layer and therefore are not vetoed. Some fraction of these photons is detected in layers 2 and 3. For these layers the contribution from these photons is not small. In fact, these L-escape photons from the top layer  contribute significantly to the overall signal in these layers which can be seen by eye as apparent spikes in spectra of layer 2 and 3 below PCA channel 10.  This effect was not accounted for in previous PCA calibration versions. It led to an artificial feature at about 4$\sim$5 keV, when data from all PCU layers were analyzed jointly, which is a common approach. In the PCARMF v11.7 The L-escape contribution is described for each layer individually. According to expectations, the normalization for L-escape line is effectively zero for the top layer and non-zero for the second and third layers. They are implemented as parameters
$EscNormL2$ and $EscNormL3$. 


 \section{Calibration data}

 The PCA calibration relies on the data from the Crab nebula and the on-board \am\ calibration source.
 Each PCU is equipped with  a radioactive \am calibration source which produces  six fixed-energy lines at 13.93, 17.53, 21.13, 26.35, 29.8, 59.54 keV. 
The Crab  provides information on quantum efficiency, while the \am\ data constrain the energy redistribution
 and e2c relationship. The PCA "calibration" data set is a collection
of spectra for selected dates (given in Table \ref{caldates}), quasi-uniformly covering an entire RXTE mission span,
 beginning on April 15, 1996 (MJD 50188), i.e. the start time of the calibration epoch 3.
Observations taken before this date during epoch 1 and 2, i.e. during the first few months of 
\xte\ in-orbit performance, are regarded as a 
science validation and verification observations. 
Moreover, high voltage setting during this period may have resulted in 
non-linear effects which are not accounted for in the physical PCA response model.
This can result in a systematic effects which would affect the response quality for the entire mission.
To avoid this, we excluded observations taken before April 15, 1996 from the "calibration" data.

The PCA calibration observations are collected in two different data modes. The \am\ data are available in GoodXenon PCA data mode, providing the most detailed event description in 256 energy channels. The Crab observations are packaged in Standard2 mode, having 129 energy channels. 
We implemented two XSPEC spectral models producing 129 bin spectrum for the absorbed power law input source
spectrum and 256-bin spectrum  for the sum of six Gaussian   to model the Crab emission and the \am\ energy-to-channel
calibration source spectra correspondingly.

Calibration spectra were extracted using the following strategy.
We first selected a sample of  20 longest pointed \xte\ observations of the Crab roughly uniformly
covering the period between 1996 and 2012 for which all five PCUs were active.
Although the most clean \am\ data are available during the dedicated background observations,
we found that using individual observations does not provide enough statistics to fit 
individual \am\ spectral lines. We, therefore, have taken the following path to generate \am\
calibration spectra. We identified the dates when optimal number of the 
background observations are accompanied by the 
observations of faint sources, for which the \am\ calibration 
data is not strongly  contaminated by the signal from the observed source.
For each selected date we generated good time intervals for an entire
day by filtering out the periods when the total PCU count rate exceeds
1500 counts per second. This would exclude observations of very bright sources
such as Sco X-1, GRS 1915+105, Cyg X-1. Using this selection criteria, 
 spectra with exposure of a few tens of kiloseconds 
were produced for individual dates, so the evolution of the energy-to-channel
relationship can be modeled. With these selections in place we extracted
spectra for each individual layer of each PCUs, applying additional 
standard selection criteria to filter out episodes of Earth occultations,
South Atlantic Anomaly passages, PCU breakdowns, etc.
Examples of the Crab nebula and \am\ calibration spectra
are shown in Figure \ref{data}. In Table \ref{caldates} we show dates when the "calibration"
observations were taken.

\section{The response minimization method}
\label{setup}

The  {\tt pcarmf} parameter minimization procedure implemented prior to the version v11.7 
was divided into several steps. First, the e2c parameters were obtained by approximating
the e2c relationship to best represent energies of the \am\ lines and the iron K$_\alpha$
lines derived from a set of Cas-A observations. Then, individual Crab observations were
fitted with the response model. Because of lack of sensitivity of the fits to individual 
observations, only a subset of response parameters was allowed to vary in this approach. 
This required several subsequent runs to optimize different parameter subsets. As the last 
step, the results were averaged to get the final parameter values.
This method provided a reliable PCA calibration
for the data taken before 2004. However, for more recent observations calibration tests showed
degrading quality of the energy-to-channel relationship as well as representation of the 
Crab nebula spectrum  by the calibration model. 
To produce a reliable response for the entire \xte\ performance period of more than 15 years, a new, more effective
method of response parameter minimization was required. 

To optimize a response modeling procedure we created 
a new \xte/PCA calibration environment within XSPEC astrophysical fitting software.
First, we implemented the PCA response model as an XSPEC model. Because the e2c-relationship
is a part of the model, it operates in a raw PCA instrumental channel space.  Namely, assuming
 a particular source spectrum (e.g. power law or sum of Gaussians), it convolves the spectral 
 shape with the PCA response, defined by a set of parameters identical to {\tt pcarmf} parameters,
 and yields the expected number of counts in each spectral channel for a given input energy spectrum. 
 The model has the same set of parameters as  {\tt pcarmf} task plus normalization, 
 which is effectively a PCU area modified
 by a PCU offset factor (see below).
 
 \begin{figure}
 \vspace{0.2in}
\includegraphics[scale=0.37,angle=-90]{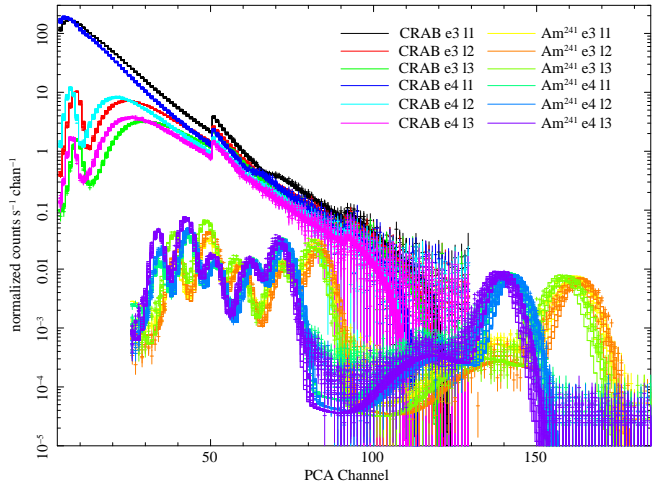}
\caption{PCA response fit to the PCU 2 data in XSPEC.}
\label{pcu2session}
\end{figure}

A set of PCA response parameters describing one particular PCU unit is obtained by fitting 
a "calibration" set of the Crab and \am\ spectra. 

After data selection and extraction
  "calibration" spectra are loaded in one XSPEC session and a separate response model
is assigned to each spectrum, convolved with an appropriate source spectrum (e.g.
absorbed power law for Crab or a sum of six Gaussians for the \am\ source).
In the Crab spectra we ignore PCA Standard2 mode channels 1-3, which corresponds to
the first seven channels of the raw 
PCU 256 channels.
The \am\ data below PCU channel 25 do not contain any information on the 
e2c relationship and are also ignored in our fits.
Despite the enormous number of total parameters for all model components (several 
thousands per fitting session), the number of independent parameters is less then a hundred
as most parameters are interlinked with each other.

The following parameter linking has been implemented:
\begin{itemize}
\item all spectra have their source spectral parameters
(i.e. power law index and normalization for the Crab spectra and
parameters describing six gaussians for the \am\ data) interlinked and fixed, except for the Gaussian
normalizations which are free but interlinked for
spectra from the same layer
\item all spectra had 
the parameters describing xenon and propane amounts
and parameters describing PCU geometry (i.e. normalization,
thickness of mylar and aluminum windows)
linked for all spectra  
\item parameters of physical model describing quantum
efficiency and redistribution are the same for all "calibration spectra"
\item energy-to-channel parameters are the same for all
spectra belonging to the same layer and gain epoch
\end{itemize}

As a spectral model for the Crab emission spectra we used a standard absorbed 
power law model with the same Crab spectral properties 
which were assumed for previous PCA response versions:
the index $\Gamma_{Crab}= 2.11$, normalization $N_{Crab}=11.0$ 
and $N_{H\ Crab} = 0.34 \times 10^{22}$ cm$^{-2}$. 
Recent analyses of X-ray multi-mission data show evidence 
for the departure of the Crab spectral shape from a simple power 
in a broad band spectral range  \citet{kirsch,wei10}. However,
there is no reliable alternative beyond this model has been identified yet.
We therefore tentatively follow the \citet{wei10} conclusion that the approach where response 
parameters are tuned to approximate a smooth shape for the Crab 
spectrum is justified. The PCA response minimization procedure described
above can be easily modified to accommodate any spectral form
of the calibration target. It will be straightforward to 
recalibrate the PCA to a new shape of the Crab spectrum, when
it becomes available.

 We note, that 
in recent analysis of multi-mission lightcurve from the  Crab nebula (including major contribution from
\xte )  \citet{wh11} identified a quasi-periodic modulation in the Crab flux with a period
of approximately 3 year and amplitude of a several percent. Our analysis of the Crab pulsed emission, also reported 
in \citet{wh11}, have strongly suggested the nebula as a origin of the modulation.
This result would undermine validity of the calibration obtained by fitting the Crab data
collected over periods much smaller than the observed periodicity.
However, we model the \xte/PCA response based on the data collected 
for $\sim$ 15 years. We, therefore, assume that any variations in
Crab spectrum are averaged out. In \S 5 we provide a check
for this assumption by fitting the absorbed power law to the 
individual Crab observations throughout the entire mission span. The
power law parameters show small, mostly short-term variations around the
calibration values, which are expected by have a small effect on the
overall fit.

As a first step, we determined 
response parameters for PCU 2 we used data for 20 Crab 
observations and \am\ spectra for 15 days totaling in 105
individual spectra. The model have 5445 parameters in total
from which 74 are free. After a minimum fit statistic
is achieved for the session including both Crab and 
\am\ data, we perform an additional fit on a reduced
data set using the Crab data only with the e2c coefficients
fixed. This is done to remove any possible contribution
from the unmodeled residuals from \am\ data to the response
parameter values.  The resulting parameter values are given in 
Tables \ref{pcu2pars} and \ref{pcu2e2c}. Then, this procedure is repeated for other PCUs, although
this time we fixed the PCA parameters universal for all PCUs, i.e.
$kEdge\_veto$,$lEdge\_veto$, etc., (see Table \ref{pcu2pars}). This implies
that the universal PCA parameters obtained by the fit to PCU2
data are valid for all other PCUs. This is a good assumption 
as the PCA universal parameters describe PCU physics and
geometry which are quite similar for all PCUs.

\section{Results}

As a zeroth approximation for our fits we use have used 
parameter values from the previous release of
PCA calibration, i.e. PCARMF v11.1 and e2c {\it e05v03}.
We remind that the previous e2c table
has 5 epochs, with the fifth epoch starting 
on May 13, 2000 00:00 (MJD 51677). As described below,
in the new e2c table the fifth epoch is retained for 
PCUs 0 and 1 only with individual start dates corresponding to
the moments of a propane layer loss. 

Major improvements in performance was achieved
as a result of the following modifications 
to the PCA response model and e2c relationship:
\begin{itemize}
\item First, we observed that the non-linear terms of the time dependence in the e2c relationship,
namely, coefficients $A_2$ and $B_2$ in Equations 4 and 5 of J06
are not required to account for  the e2c relationship evolution .
This is consistent with a slow linear time evolution as a result 
of a small leak of xenon from the first layer to the propane veto layer.
Fixing these parameters at
zero led to a dramatic improvement in the convergence speed and fit quality.
Subsequent thawing of these parameters did not improve the statistic and 
showed that their values are consistent with being effectively zero. 

\begin{figure}
\vspace{0.2in}
\includegraphics[scale=0.85,angle=-90]{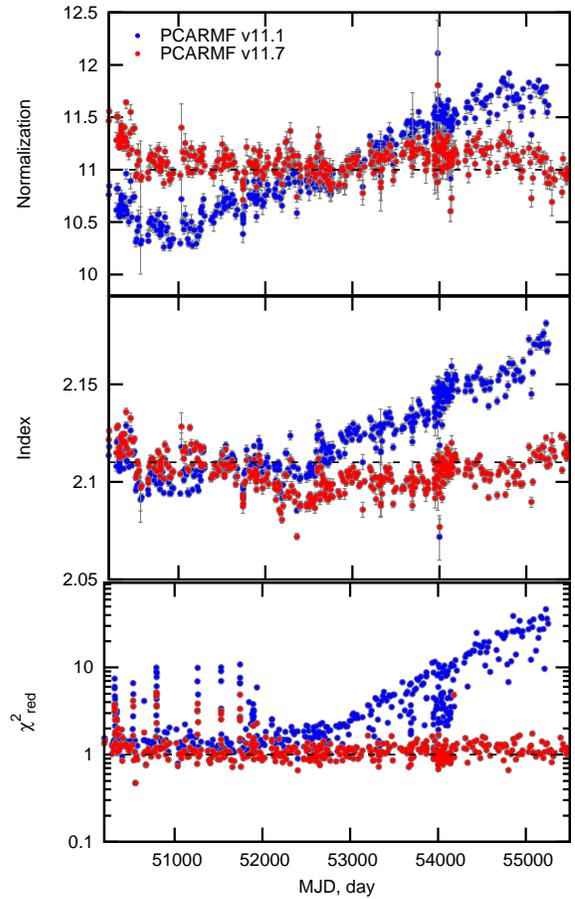}
\caption{The Index (top), normalization (middle) and the reduced $\chi^2$ (bottom) of the absorbed power law fit to the Crab nebula {\it RXTE} spectra throughout the mission. The spectra from the top layer of the PCU 2 are used for fits.  Blue data points show fit results  using the PCARMF v11.1, while red points show results of the new response (PCARMF v11.7). }
\label{crab_test}
\end{figure}

\item The next important
observation concerning the behavior of e2c relationship
was that the e2c parameters obtained for epoch 5 were statistically 
identical to the best-fit values  for corresponding parameters of e2c  epoch 4.
This effectively showed that, in accordance with expectations, 
a new e2c epoch is required only when an abrupt change in e2c relationship
occurs either due to the PCU anode voltage change or due to
the loss of the PCU propane layer. This renders the epoch 5 obsolete
for PCUs 2,3 and 4. Epoch 5 is retained for PCUs 0 and 1 starting at the
corresponding date of propane layer loss.
\item The energy resolution in channel space is modeled as 
\begin{equation}
\Delta ch = (\sqrt{aE+b})B
\end{equation}
where B is defined in Equation \ref{b}. In e2c $e05v03$ energy resolution coefficients are set to
$a=0.121$ and $b=0.442$ to formally satisfy the ground test which showed resolution $\Delta E/E \sim 0.17$
at 6 keV and $\sim 0.08$ at 22 keV (J06). We note, however, that two values are generally 
not constrained by two measurements. Our fits to the \am\ data showed that resolution coefficients have quite different values, i.e. $a\approx0.18$ and
$b\approx0.0$ (see below), which are dictated by \am\ line widths and are also consistent with the
ground prelaunch data. 
\end{itemize}

We show PCU 2 calibration spectra fitted with the PCA response model
in Figure \ref{pcu2session}.  The final set of best fit parameters is given
in Table \ref{pcu2pars}. In Table \ref{pcu2e2c} we also present
a complete set of e2c parameters. The quality of the model fit
which includes   the Crab data only is $\chi^2/N_{dof}= 3.4$. 

This quite impressive, taking the fact that 
 we did not assume any systematic error in the data.
 In light of the evidence of the multi-year periodicity in the Crab
 emission at a level of a few per cent presented by \citet{wh11}, which is not modeled in our fits,
  the achieved fit quality may indicate that we are nearing the statistical limit and that the
  further attempts to improve the fit will not lead to actual improvement of the PCA instrument response.
  This however, allow us to estimate the range of systematic error to use in spectral fitting of \xte/PCA 
  spectra. Namely, if we assume  systematic error of 0.5\%,1\% and 1.5\% we get  the corresponding fit quality of  1.9 and
 1.6 and 1.3. This illustrates the range of systematic errors to use for \xte\ spectral analysis. 
For the most observations, the systematic error of 0.5\% is sufficient, while for the extreme cases it can be 
raised up to 1.5\%. 

The response model fits quality achieved for the PCUs 3 and 4 are similar to 
that achieved for the PCU2 indicating
that the determining the PCA universal response parameters 
using the PCU2 data only is justified. The data for the
PCUs 0 and 1 after the propane layers loss events 
are contaminated by particle flux and therefore show slightly
worse fit quality as expected. For reliable spectral 
analysis it is recommended to exclude 
the data from PCUs 0 and 1 after the propane 
layer loss.

\label{disc}
To check the consistency and the quality of the resulting response 
we fitted the complete set of {\it RXTE} observations of Crab throughout
the mission with the absorbed power law model keeping the 
$N_H$ frozen at $0.34\times10^{22}$ cm$^{-3}$ and keeping the index and normalization
 free. We performed the fits with both the previous and the new response versions.
We show the results of the test fits of the Crab nebula data from PCU 2 in Figure \ref{crab_test}.
It can be clearly seen that the quality of the PCA calibration provided by 
 the PCARMF v11.1 is degrading exponentially starting around MJD 52500.
 On the other hand  PCARMF v11.7 combined with e2c {\it e03v05} are showing
 uniform fit quality with the reduced $\chi^2\sim$ 1.0 throughout the entire
 period of \xte\ performance.  The variations in the power law index and normalization 
 seen in Figure \ref{crab_test} are consistent with the variations in the
 count rates from the Crab reported in \citet{wh11}. While in the power law fit parameters 
 the variations are less visible,  a closer examination of their
 time dependence beginning from MJD 52000 (the start date of the \citet{wh11} analysis), 
 especially the index, reveals the correlation. Namely, the index shows the upward
 trend along with superimposed variations with peaks at MJD 53000 and 54000 in agreement
 with the downward trend and minima around the same dates seen in the 
 lightcurves shown in \citet{wh11}. While the global trend for the data shown
 in Fig. \ref{crab_test} breaks around MJD 52500, the hint for the variations 
 is still present before this date in support of the main claim of the \citet{wh11}
 that the Crab nebular exhibits the long term variations. 

\begin{figure}
\includegraphics[scale=0.33,angle=-90]{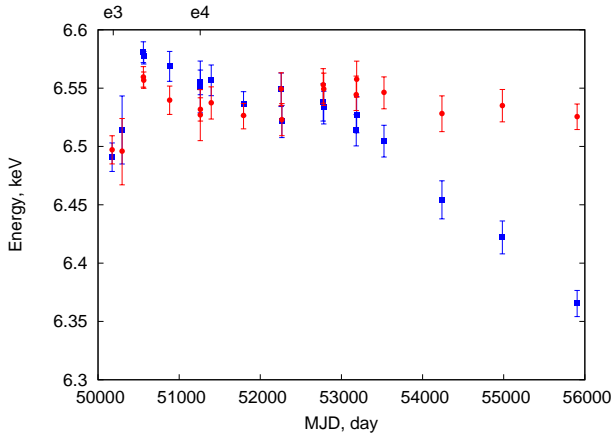}
\caption{ The Fe 6.5 keV line complex from a complete set of PCA observations of Cas A fitted with a narrow gaussian.
The data from the top layer of PCU 2 is used. The red color shows the PCARMF v11.7 results, while blue is used for the PCARMF v11.1. Labels at the top axis mark the starts of the epochs 3 and 4 respectively}
\label{casa_test}
\end{figure}

In the PCA response versions prior to the PCARMF v11.7
the data from the supernova remnant Cas A, which shows multiple emission lines
in soft X-rays \citep{holt}, was used to constrain e2c relationship by utilizing
the energy of the iron line complex at $\sim$6.5 keV. We took a slightly different approach
by not using the Cas A iron line in the response minimization. Instead, as we fit 
the Crab data letting both quantum efficiency and e2c parameters change free, we expect
that information in the Crab spectrum at the low energies due to
xenon L-edges would constrain the energy scale. The Cas A data thus provide a means to check the
 consistence of the energy scale below 10 keV. Cas A was observed by RXTE on a yearly basis.
We fit the Cas A spectrum for the PCU 2 layer 1 between 5 and 14 keV with the model
consisting of a power law and two gaussians for the emission line features at 6.5 keV and 8.1 keV. 
In Fig. \ref{casa_test} we show the results for the new and old PCARMF versions, which clearly show that
the PCARMF v11.7 e2c relationship is much more reliable giving a stable energy 
for the 6.5 keV complex line energy at about 6.54 keV. The old response shows 
a departure from the 6.5 keV region towards lower energies. 

\section{Conclusions} 
\label{summary}

We present a new \xte\/PCA response version v11.7. This response 
is based on data set presenting the entire \xte\ mission span. While the new
response is largely based on the physical model developed in J06, some 
significant modifications are made, especially for energy-to-channel conversion 
relationship. The new response shows a superior performance with respect to the
previous \xte\ response versions.

\begin{deluxetable}{lll|ll} 
\tablecolumns{3} 
\tablewidth{0pc} 
\tablecaption{PCA Calibration data set} 
\tablehead{ 
\multicolumn{3}{c}{The Crab Nebula \tablenotemark{a}}& \multicolumn{2}{c}{The \am Source}\\
\colhead{ObsID} & \colhead{Start, MJD}    &   \colhead{Exposure\tablenotemark{b} , s} & \colhead{Date, MJD} &   \colhead{ Total exposure \tablenotemark{c}, ks} \\ }

\startdata 
10202-02-01-00 & 50191.82 & 2692 & 50200 &  60.9 \\
10203-01-01-00 & 50318.01 & 9816 & 50800 &  42.0 \\
20804-01-06-00 & 50793.45 & 13031 & 51000 &  37.6 \\
30133-01-11-00 & 51021.92 & 1077 & 51210 &  35.2 \\
30133-01-18-00 & 51120.44 & 884 & 51300 &  53.0 \\
30133-01-26-00 & 51233.08 & 822 & 51810 &  60.2 \\
40093-01-03-00 & 51273.51 & 953 & 52400 &  75.2 \\
50804-01-17-00G & 51937.40 & 5753 & 53200 & 69.4   \\
60069-01-09-00 & 52273.20 & 869 & 53960 &  27.0\\
70802-01-08-00 & 52683.31 & 923 & 54531 & 29.4  \\
80802-02-09-01 & 52954.69 & 722 &  54861 &  46.85  \\
90802-02-05-00 & 53273.07 & 430 & 54991 &  37.0 \\
91802-02-01-00 & 53583.04 & 446 & 55128 &  35.07 \\
92802-03-06-00 & 54180.24 & 7745 & 55309 &35.33\\
93802-02-12-00 & 54523.10 & 936.0 & 55576 & 38.53 \\
94802-01-03-00 & 54856.40 & 1322 & & \\
94802-01-08-00 & 54929.98 & 1167 & & \\
94802-01-22-00 & 55170.75 & 922 & & \\
95802-01-03-00 & 55228.32 & 968 & & \\
95802-01-22-00 & 55548.60 & 998 & & \\

\enddata 
\tablenotetext{a}{The chosen observations of the Crab nebula have all 5 PCUs active}
\tablenotetext{b}{The Crab exposures are the same for all PCUs}
\tablenotetext{c}{The exposure is given for PCU2. The \am\ exposures vary for different PCUs and are usually shorter than for the PCU2.}
\label{caldates}
\end{deluxetable} 

\begin{deluxetable}{lllll} 
\tablecolumns{5} 
\tablewidth{0pc} 
\tabletypesize{\scriptsize} 
\tablecaption{PCARMF Parameters\tablenotemark{a} for PCU 2 } 
\tablehead{ 
\colhead{Parameter} & \colhead{Units}    &   \colhead{ Domain}   & \colhead{Value} & \colhead{Description}\\ }
\startdata 
{\it xe\_gmcm2\_l1} & gm/cm$^2$ & PCU Layer 1 & (6.921 $\pm$ 0.029)$\times 10^{-3}$ & Xenon amount  \\  
{\it xe\_gmcm2\_l3} & gm/cm$^2$ & PCU Layer 2 & (5.739 $\pm$ 0.026)$\times 10^{-3}$ & Xenon amount  \\  
{\it xe\_gmcm2\_l3} & gm/cm$^2$ & PCU Layer 3 & (5.799 $\pm$ 0.024)$\times 10^{-3}$ & Xenon amount  \\  
{\it xe\_gmcm2\_pr} & gm/cm$^2$ & PCU Propane Layer &  (1.312 $\pm$ 0.056)$\times 10^{-4}$& Xenon amount  at the reference date \\  
{\it xe\_gmcm2\_dl} & gm/cm$^2$ & PCU Dead Layer & (4.41 $\pm$ 0.27)$\times 10^{-4}$ & Xenon amount \\
{\it pr\_gmcm2} & gm/cm$^2$ & PCU Propane Layer & (2.646 $\pm$ 5.272)$\times 10^{-3}$& Propane amount \\ 
{\it my\_gmcm2} & gm/cm$^2$ & PCU & (6.893 $\pm$ 5.3)$\times 10^{-3}$& Mylar window thickness \\ 
{\it al\_gmcm2} & gm/cm$^2$ & PCU & (1.204 $\pm$ 4.23)$\times 10^{-4}$& Aluminum window thickness \\ 
{\it xe\_pr\_daily} & gm/cm$^2$/day & PCU & (3.93 $\pm$ 0.02)$\times 10^{-8}$& Xenon Leak Rate \\ 
{\it kEdge\_veto} & & PCA & 0.813 $\pm$ 0.007&  \\ 
{\it lEdge\_veto} & & PCA & 0.934 $\pm$ 0.005&  \\ 
{\it EscFracKa} &  & PCA &  0.399$\pm$ 0.004&  \\ 
{\it EscFracKb} &  & PCA & 0.298 $\pm$ 0.003 &  \\ 
{\it EscFracL2} &  & PCA Layer 2& (5.61 $\pm$ 0.60)$\times 10^{-3}$&  \\ 
{\it EscFracL3} & & PCA Layer 3& (2.028 $\pm$ 0.60)$\times 10^{-2}$&  \\ 
{\it EscNormKb} &  & PCA & 0.404 $\pm$ 0.022&  \\ 
{\it EscNormL2} &  & PCA Layer 2 & 5.44 $\pm$ 0.62&  \\ 
{\it EscNormL3} &  & PCA Layer 3 & 126.6 $\pm$ 12.6&  \\ 
{\it epoint}&&PCA&18.4$\pm$&\\
{\it track\_coeff}&&PCA&(3.72$\pm$)$\times 10^{-2}$&\\
{\it track\_exp}&&PCA&3.32$\pm$0.04&\\
{\it pcc\_coeff}&&PCA&(1.37$\pm$0.06)$\times 10^{-2}$&\\
{\it wxef}&&PCA&0.492$\pm$0.007&\\
{\it resp1}&&PCU&0.1733$\pm$0.0003&\\
\enddata
\tablenotetext{a}{See J06 for the detailed description and definition of individual PCA response  parameters}
\label{pcu2pars}
\end{deluxetable} 

\begin{deluxetable}{lll} 
\tablecolumns{3} 
\tablewidth{0pc} 
\tablecaption{E2C Parameters for PCU 2 } 
\tablehead{ 
\colhead{Parameter} & \colhead{Epoch 3}    &   \colhead{Epoch 4}  \\ 
 & 04/15/96-03/22/99    &   03/22/99-Present  \\ }
\startdata 
\multicolumn{3}{c}{Layer 1}\\
\cline{1-3}  \\ 
A$_0$  & -0.761$\pm$0.009 & -0.658$\pm$0.007  \\  
A$_1$  & (-5.33$\pm$0.52)$\times 10^{-5}$&(-9.12$\pm$1.22)$\times 10^{-6}$\\  
B$_0$  & 2.895$\pm$0.001&  2.476$\pm$0.001   \\  
B$_1$  & (6.22$\pm$0.08)$\times 10^{-5}$ & (1.05$\pm$0.01)$\times 10^{-5}$  \\  
C$_0$  & (-6.84$\pm$0.06)$\times 10^{-3}$ & (-5.17$\pm$0.04)$\times 10^{-3}$  \\  
D$_0$  & (5.57$\pm$0.09)$\times 10^{-5}$ & (4.10$\pm$0.06)$\times 10^{-5}$  \\  
\cline{1-3}  \\ 
\multicolumn{3}{c}{Layer 2}\\
\cline{1-3}  \\ 
A$_0$  & -0.560$\pm$0.005 & -0.577$\pm$0.005  \\  
A$_1$  & (2.61$\pm$0.78)$\times 10^{-5}$&(-1.8$\pm$1.5)$\times 10^{-6}$\\  
B$_0$  & 2.815$\pm$0.001&  2.434$\pm$0.001   \\  
B$_1$  & (4.69$\pm$0.10)$\times 10^{-5}$ & (1.04$\pm$0.01)$\times 10^{-5}$  \\  
C$_0$  & (-1.32$\pm$0.06)$\times 10^{-3}$ & (-1.71$\pm$0.04)$\times 10^{-3}$  \\  
D$_0$  & (-7.68$\pm$0.94)$\times 10^{-6}$ & (6.9$\pm$5.5)$\times 10^{-7}$  \\  
\cline{1-3}  \\ 
\multicolumn{3}{c}{Layer 3}\\
\cline{1-3}  \\ 
A$_0$  & -0.374$\pm$0.008 & -0.217$\pm$0.008  \\  
A$_1$  & (2.7$\pm$1.7)$\times 10^{-5}$&(-5.35$\pm$3.25)$\times 10^{-6}$\\  
B$_0$  & 2.810$\pm$0.001&  2.391$\pm$0.001   \\  
B$_1$  & (4.35$\pm$0.16)$\times 10^{-5}$ & (1.00$\pm$0.02)$\times 10^{-5}$  \\  
C$_0$  & (-5.45$\pm$0.06)$\times 10^{-3}$ & (-3.46$\pm$0.04)$\times 10^{-3}$  \\  
D$_0$  & (4.24$\pm$0.09)$\times 10^{-5}$ & (2.46$\pm$0.05)$\times 10^{-5}$  \\  
\enddata 
\label{pcu2e2c}
\end{deluxetable}



\end{document}